\documentclass[reprint,showpacs,preprintnumbers,amsmath,amssymb, citeautoscript,prl,superscriptaddress]{revtex4-1}
\usepackage{graphicx,wasysym}
\usepackage{dcolumn}
\usepackage{bm}
\setcitestyle{super}
\usepackage{natbib}
\usepackage{tabulary}
\usepackage{hyperref}

\newcommand{\tN}{t_{\mathrm{N}}}

\newcommand{\kB}{k_\mathrm{B}}

\usepackage{bbold}

\begin{document}

\preprint{1}

\title{Heat current fluctuations and anomalous transport \\ in low dimensional carbon lattices}
\author{Ushnish Ray}
\affiliation{%
Department of Chemistry, California Institute of Technology, Pasadena, CA 08540
}%
\author{David T. Limmer}
 \email{dlimmer@berkeley.edu}
\affiliation{%
Department of Chemistry, University of California, Berkeley, CA 94609
}%
\affiliation{%
Kavli Energy NanoScience Institute, Berkeley, CA 94609
}%
\affiliation{%
Chemical Science Division, Lawrence Berkeley National Laboratory, Berkeley, CA 94609
}%
\affiliation{%
Materials Science Division, Lawrence Berkeley National Laboratory, Berkeley, CA 94609
}%

\date{\today}
\begin{abstract}
Molecular dynamics simulations and methods of importance sampling are used to study the heat transport of low dimensional carbon lattices. For both carbon nanotubes and graphene sheets, heat transport is found to be anomalous, violating Fourier's law of conduction with a system size dependent thermal conductivity and concomitant nonlinear temperature profiles. For carbon nanotubes, the thermal conductivity is found to increase as the square root of the length of the nanotube, while for graphene sheets the thermal conductivity  is found to increase as the logarithm of the length of the sheet over the system sizes considered. The particular length dependence and nonlinear temperature profiles place carbon lattices into a universality class with nonlinear lattice models, and suggest that heat transport through carbon nano-structures is better described by a Levy walk rather than simple diffusion.
\end{abstract}

\maketitle

The heat transport properties of low dimensional lattices have received considerable recent attention, due to experimental and simulation reports claiming a violation of Fourier's law of conduction\cite{chang2008breakdown,xu2014length,yang2010violation,wang2011non}. Experimentally, reports on both carbon nanotubes and graphene sheets have shown indications of anomalous conductivities, though difficulties extracting definitive values are complicated by boundary effects.
To date there is little theoretical consensus on the underlying mechanism or its generality to guide interpretation of such measurements\cite{lepri2016heat,dhar2008heat}.
Carbon nano-structures offer an ideal material to test predicted theoretical scaling relations, but to do so computationally requires robust simulation methods, which have thus far led to 
many contradictory reports\cite{che2000thermal,maruyama2002molecular,mingo2005length,wang2006carbon,cao2012size,nika2012anomalous,pereira2013divergence,park2013length}.

Here we use molecular simulation and dynamic importance sampling\cite{ray2018importance,gao2017transport,gao2018nonlinear} to show that thermal transport in carbon nanotubes and graphene sheets violates Fourier's law. Specifically, we demonstrate that over the system sizes considered the 
 thermal conductivity of low dimensional carbon lattices increases with their characteristic size, in a manner dependent on dimensionality and  distinct from finite size effects related to phonon confinement.
Our results clarify that heat transport in low dimensional carbon lattices is anomalous despite the statistics of heat currents being Gaussian and linear response remaining valid. Rather, the size dependent thermal conductivity is due to the confinement of momentum fluctuations, resulting in slowly decaying heat current correlations. These correlations result in energy transport that is better described by a Levy walk, rather than a simple diffusive process, as has been proposed for simple nonlinear lattice models\cite{denisov2003dynamical,cipriani2005anomalous}.

In macroscopically three-dimensional materials, Fourier's law, $j = - \kappa \nabla T$, connects a heat current, $j$, to a temperature gradient, $\nabla T$, through a material dependent constant, the thermal conductivity, $\kappa$. In low dimensional systems, those whose extents can be taken arbitrarily large in only one or two spatial dimensions, and which conserve momentum, the heat current for fixed  boundary temperatures has been found to scale with the characteristic size of the system, $L$, as $j \sim L^{-1+\alpha}$, where $\alpha$ is an anomalous exponent between 0 and 1\cite{lepri1997heat}. This has been interpreted as a size dependent conductivity, $\kappa_L \sim L^{\alpha}$, that diverges in the thermodynamic limit of $L \rightarrow \infty$ in analogy with other examples of long-time tail behavior\cite{berne1971topics}. Both simulations and theory for simple nonlinear lattices agree that this divergence occurs, though the value of $\alpha$ and the underlying mechanism are debated. Recent mode-coupling theory\cite{delfini2007anomalous} and renormalization group calculations\cite{spohn2014nonlinear} predict two distinct universality classes with $\alpha=1/3$ or 1/2, for 1$d$ systems depending on the dominant nonlinearity and boundary condition\cite{lee2015universality}, but both expect a logarithmic divergence for 2$d$ systems\cite{lepri2016heat}.   
These findings are supported by some numerical calculations, though others have reported distinct exponents\cite{mai2007equilibration,delfini2008comment}, as well as a sensitivity of $\alpha$ to model details\cite{hurtado2016violation}. 

Simulation studies on carbon nanostructures are more limited, and at present there is not agreement on whether transport is normal or anomalous. Studies on nanotubes have reported normal transport\cite{che2000thermal,maruyama2002molecular,mingo2005length}, though others have reported exponents of $\alpha=1/4$ to $\alpha=1/2$ \cite{yao2005thermal,wang2006carbon,shiomi2008molecular}. It has been argued that out of plane flexural modes in $2d$ graphene sheets tame a logarithmic divergence\cite{nika2012anomalous}, though if sufficiently strained, anomalous transport is claimed to be restored\cite{pereira2013divergence}.  Most simulations reporting anomalous transport employ nonequilibrium simulation techniques, which are typically faster to converge for large systems than analogous equilibrium calculations based on Green-Kubo theory. This has led researchers to question whether anomalous transport  in realistic lattices a nonequilibrium effect.  

In order to determine which, if either, universality class carbon lattices fall into, we consider both a single walled carbon nanotube and a graphene sheet. In both cases, we consider only heat transported from classical nuclear degrees of freedom. To simulate these systems, we embed the isotopically pure solids in periodic boundary conditions, and orient them such that the largest length, $2L+2\delta$, is along the $z$ direction. The size of the simulation box is set to ensure no residual lattice strain. The individual atoms evolve through the equation of motion,
\begin{eqnarray}
\label{Eq:Lang}
m \dot{\mathbf{v}}_i(t) &=&\mathbf{F}_i[\mathbf{x}(t)] - \gamma_i \mathbf{v}_i(t) + \mathbf{R}_i(t) 
\end{eqnarray}
where $m$ is the mass of a carbon atom, $\mathbf{v}_i$ is the $i$th atom's velocity, $\mathbf{F}_i[\mathbf{x}(t)]$ is the conservative force acting on the $i$th atom from the other atoms, described here by the gradient of a Tersoff potential parameterized to recover the phonon spectrum of carbon nanostructures\cite{lindsay2010optimized}. The second and third terms in Eq.~\ref{Eq:Lang} describe a Langevin thermostat that obeys a local detailed balance
 with a temperature $T_i$, by dissipating energy through the friction, $\gamma_i$, and adding energy by a random force $\mathbf{R}_i(t)$ with Gaussian statistics described by $\langle \mathbf{R}_i(t) \rangle = 0$, $\langle \mathbf{R}_i(t)\mathbf{R}^{T}_j(t') \rangle = 2 \gamma_i \kB T_i \delta_{ij}\delta(t-t')\mathbb{1}$, where $\kB$ is Boltzmann's constant. 
 
In all simulations, two distinct thermostats act on groups of atoms, denoted by $l$ and $r$, each over a region of length $\delta$ along the $z$ direction and are separated by a distance $L$. In these two regions, $m/\gamma_i=1$ ps and the temperature in each group is set to $T_{l/r} = T\pm \Delta T/2$. Outside of these thermostat regions, $\gamma_i = 0$, and the atoms evolve with Hamiltonian dynamics. For all calculations we consider $T=300$ K, $\delta = 1$ nm, and have checked that our choice of $\delta$ is large enough given $\gamma$ to not affect our observed scaling behavior, as consistent with other observations.\cite{salaway2014molecular} We consider a (10,10) nanotube, and a graphene sheet with width 20 nm, which is sufficient to converge effects from finite width.

The heat transport through the carbon lattices is studied by monitoring the energy exchanged with the stochastic thermostats that act as ideal reservoirs. Specifically, the energy current through the $k$th reservoir is given by a sum over $N_k$ atoms in that region,
\begin{equation}
\label{Eq:jk}
j_k(t) = \sum_{i \in k}^{N_k} \left [- \gamma_i \mathbf{v}_i(t) + \mathbf{R}_i(t)\right ]\cdot \mathbf{v}_i(t)
\end{equation}
and thus the energy exchanged from the $r$th reservoir into the $l$th reservoir over a time $t_\mathrm{N}$ is the integrated current 
\begin{equation}
J(t_\mathrm{N}) =  \int_0^{t_\mathrm{N}} dt\,[ j_l(t)-j_r(t) ]
\end{equation}
where Eq.~\ref{Eq:jk} is interpreted in the Stratonovich sense. If the system is driven into a nonequilibrium steady-state by maintaining a temperature difference between the two reservoirs, the thermal conductivity can be defined as, 
\begin{equation}
\label{Eq:KapNEq}
\kappa_L = \lim_{\Delta T \rightarrow 0} \lim_{t_\mathrm{N} \rightarrow \infty} -\frac{\langle J(t_\mathrm{N}) \rangle_{\Delta T}}{t_\mathrm{N} \Delta T}L
\end{equation}
where the long time limit is taken to ensure a steady-state, $\Delta T$ is taken small as the  conductivity is defined as a linear response coefficient, and $\langle \dots \rangle_{\Delta T}$ denotes a stochastic average at fixed $\Delta T$. If Fourier's law holds, $\Delta T/L$ can be identified as the temperature gradient in the limit that $L\rightarrow \infty$.

Alternatively, if the system is maintained at thermal equilibrium, where the reservoirs are fixed to a common temperature, $T$, the conductivity is computable from the mean-squared fluctuations of the total energy exchanged,
\begin{equation}
\label{Eq:KapEq}
\kappa_L = \lim_{t_\mathrm{N} \rightarrow \infty} \frac{\langle  J^2(t_\mathrm{N}) \rangle_{0}}{2 t_\mathrm{N} \kB T^2}L
\end{equation}
where at long times, for a finite open system, the mean-squared fluctuations are expected to scale linearly with time. This exact expression follows from the definition of $\kappa_L$ in Eq.~\ref{Eq:KapNEq} and the stochastic process in Eq.~\ref{Eq:Lang}, and is an example of an Einstein-Helfand moment, equivalent to a Green-Kubo relation\cite{viscardy2007transport}. These two expressions for $\kappa_L$ offer independent means for studying the system size dependence of the thermal conductivity, and their equivalency reports on the domain of validity of linear response, and the ergodicity of the lattices considered. 

\begin{figure*}[t]
\begin{center}
\includegraphics[width=17.cm]{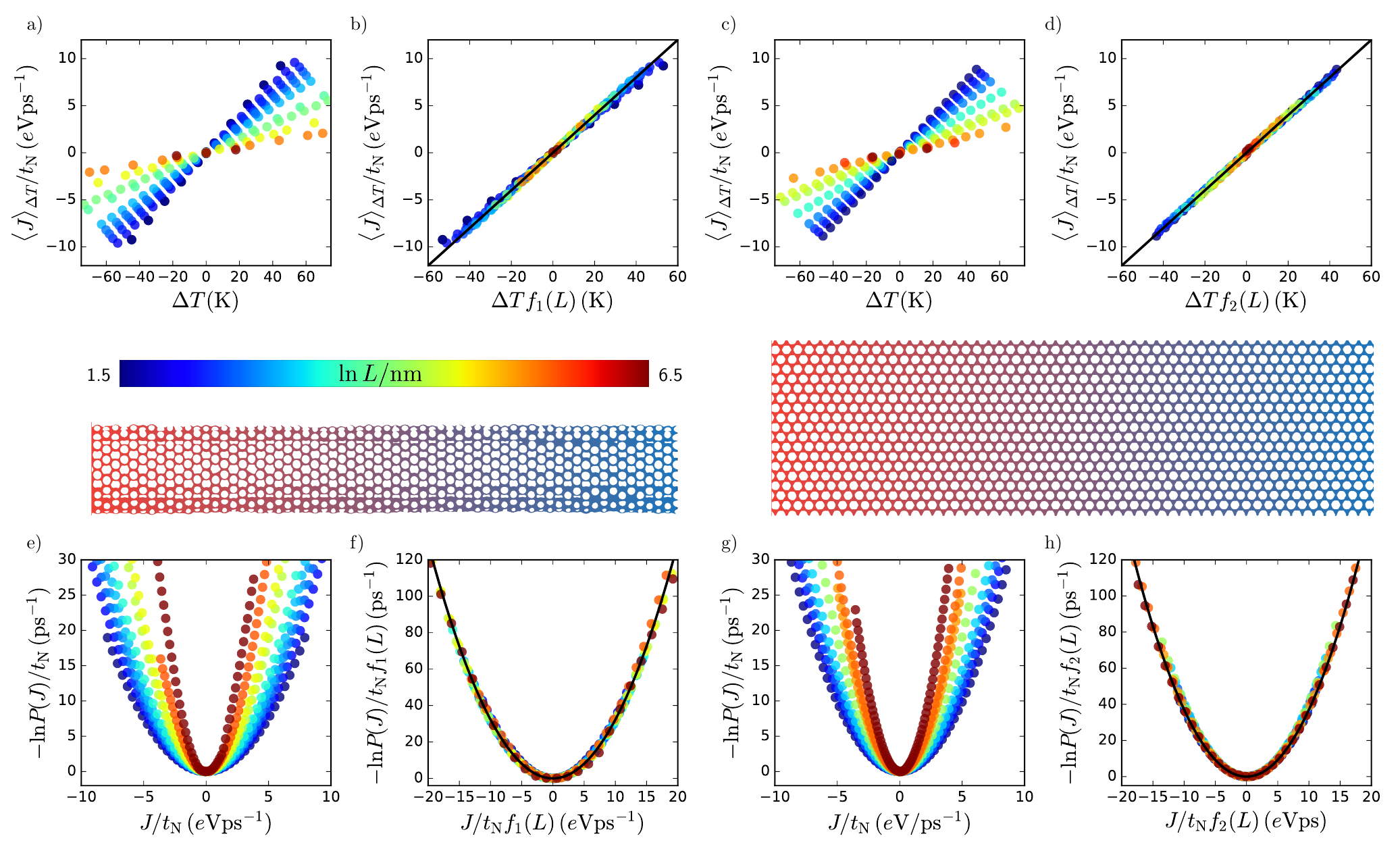}
\caption{Fluctuations and response of the integrated heat current. Mean integrated current for a) carbon nanotubes and c) graphene sheets as a function of boundary temperature difference $\Delta T$. Scaled integrated current for b) carbon nanotubes and d) graphene sheets. Probability of the integrated current for e) carbon nanotubes and g) graphene sheets at equilibrium over $\tN$. Scaled integrated current distributions for f) carbon nanotubes and h) graphene sheets. The color bar indicates the system size, $L$, throughout (a-h), the solid lines in (b,f) are guides to the eye, and the insets show typical snapshots of the systems.}
\label{Fi:1}
\end{center} 
\end{figure*}

Shown in Figs.~\ref{Fi:1}a,c) are the results of the nonequilibrium response of the integrated current to an imposed temperature bias, using $\tN=10$ ns, for both carbon nanotubes and the graphene sheets for a variety of $L$'s that span 1 nm to 1 $\mu$m as a function of the boundary temperature difference, $\Delta T$. For the nanotubes and the sheets, of all lengths considered and temperature differences up to $\pm$ 70 K, we find a linear relationship between the integrated current and the thermodynamic bias. The linearity implies that the conductivity can be extracted through Eq.~\ref{Eq:KapNEq}. If Fourier's law were valid, we would expect to find the slope of $J$ versus $\Delta T$ decrease linearly with $L$.  Instead, the data are collapsed using a scaling form $a f_{d}(L) =\langle J \rangle_{\Delta T}/t_\mathrm{N} \Delta T$ as shown in Figs.~\ref{Fi:1}b,d). The dimensionless scaling function, $f_{d}(L)$, depends on the spatial dimension $d$, where 
\begin{equation}
f_{d}(L) =\begin{cases}
\left (1+\sqrt{L/\ell_1} \right )^{-1} \, ,\quad d=1\\
\left (1+L/\ell_2 \ln{L/ \mathrm{nm}} \right )^{-1} \, , \quad d=2\\
\end{cases}
\end{equation}
interpolates between a ballistic limit for small $L$ where $\langle J \rangle_{\Delta T}$ is independent of $L$ and an anomalous limit for large $L$ where for $d=1$, $\langle J \rangle_{\Delta T} \sim L^{-1/2}$, implying that $\alpha=1/2$ and for $d=2$, $\langle J \rangle_{\Delta T} \sim \ln L /L$. The lengths, $\ell_1=3.1$ nm and $\ell_2=3.4$ nm, and $a$ is a constant equal to 1 m$e$V/ps K per carbon atom in the reservoirs.

Studying heat transport from the corresponding equilibrium fluctuations is cumbersome because it requires converging a second moment and thus necessitates long averaging times\cite{jones2012adaptive}. This has lead to a dearth of reports employing equilibrium calculations in these systems for large $L$. In order to achieve high statistical accuracy, we employ a recently developed importance sampling scheme\cite{gao2017transport}. Rather than targeting the second moment directly, we importance sample the probability of a given fluctuation in $J$, $P(J)$, using a variant of diffusion Monte Carlo for nonequilibrium steady-states known as the cloning algorithm\cite{giardina2006direct}. Specifically, we sample tilted, or deformed, distributions of the exchanged energy,
\begin{equation}
\label{Eq:KapEq}
P_\lambda(J) =  P(J)e^{ - \lambda J + \psi(\lambda) \tN}
\end{equation}
where $\lambda$ is a statistical biasing parameter that reweights fluctuations in $J$, and $\psi(\lambda) \tN= -\ln \langle \exp[- \lambda J ] \rangle_0$ is a scaled cumulant generating function that normalizes the new distribution. Using a range of $\lambda$'s, we can relate a set of $P_\lambda(J)$ to $P(J)$ using histogram reweighting techniques\cite{frenkel2001understanding,ray2018importance}, enabling us to construct $P(J)$ far into the tails of the distribution. From the distribution, we can arrive at a statistically superior estimate of linear and nonlinear transport coefficients, over Green-Kubo calculations\cite{gao2017transport,gao2018nonlinear}.

\begin{figure}[t]
\begin{center}
\includegraphics[width=8.5cm]{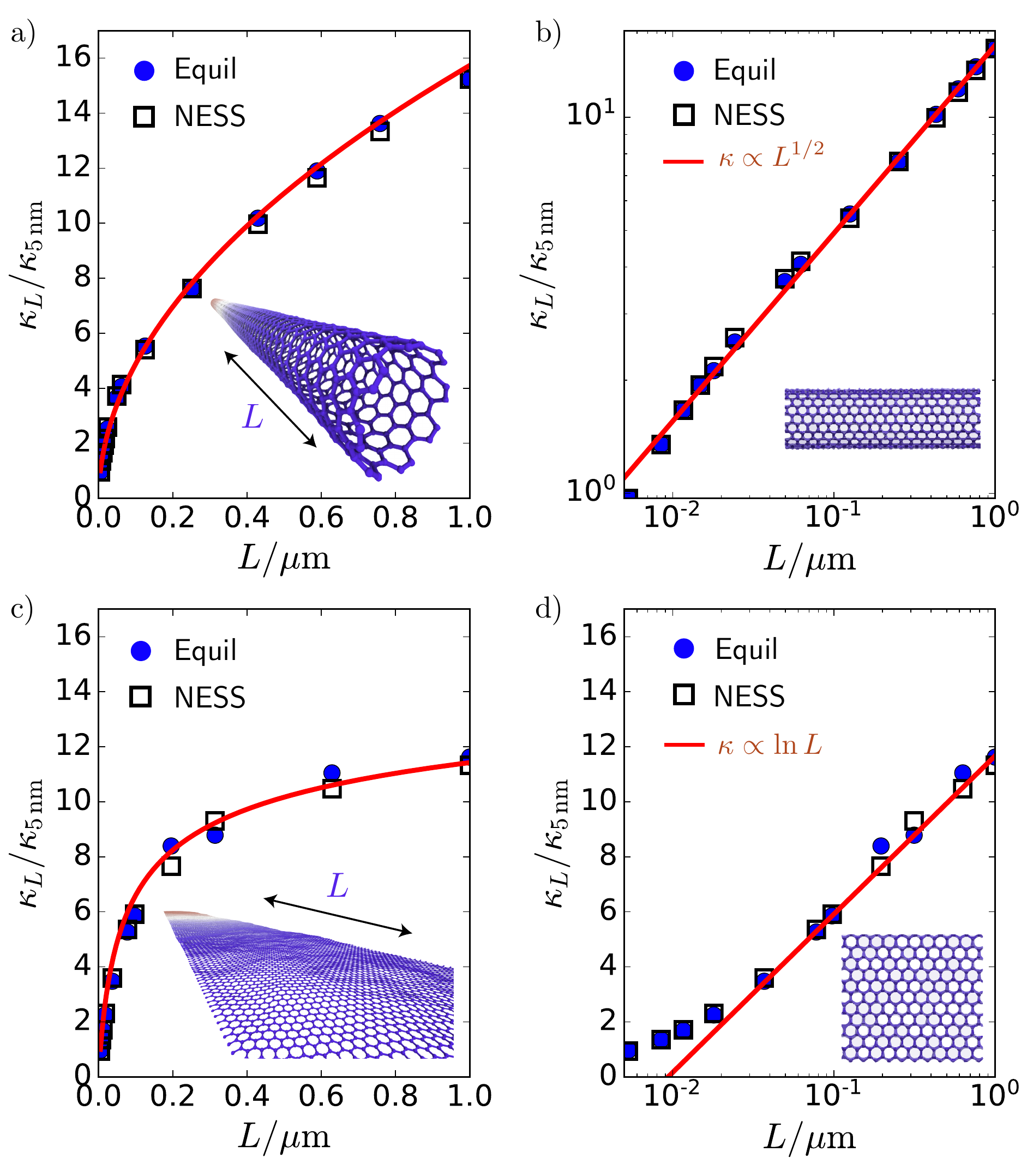}
\caption{Size dependent conductivities for a,b) carbon nanotubes and c,d) graphene sheets. In a,c) the red lines are proportional to $f_d(L)L$ and in b,d) they are guides to the eye. Filled symbols are computed from equilibrium fluctuations using Eq.~\ref{Eq:KapEq} and empty symbols are computed from nonequilbrium steady states using Eq.~\ref{Eq:KapNEq} Errorbars computed from one standard deviation are the size of the symbols.}
\label{Fi:2}
\end{center} 
\end{figure}

Shown in Figs.~\ref{Fi:1}e,g) are the results of $P(J)$ for both the carbon nanotubes and the graphene sheets for a range of $L$'s. For these calculations, we use $\tN=$20 ps, a range of 10 $\lambda$'s distributed around 0, and generate $5\times 10^4$ independent trajectories to converge the results. We use the multistate Bennet acceptance ratio to combine histograms at different $\lambda$'s~\cite{shirts2008statistically}. For the range of probabilities probed, we find that the distributions are Gaussian.
From Eq.~\ref{Eq:KapEq}, the curvature of these distributions determine $\kappa_L$, and we find we can use the same scaling function, $f_{d}(L)$, to collapse the distributions. Specifically, we use a large deviation scaling form, where $J$ and $\ln P(J)$ are divided by $t_\mathrm{N}$ and $f_d(L)$, implying all moments of $J$ scale proportional to $t_\mathrm{N}$ and $f_d(L)$. This collapse is shown in Figs.~\ref{Fi:1}f,h), and uses the same $\ell_d$ as in the nonequilibrium calculations. This consistency between nonequilibrium and equilibrium calculations is a consequence of linear response, which holds despite the heat transport being anomalous. This rules out claims that the anomalous transport is a nonequilibrium effect. The size dependence of the fluctuations of the integrated current manifest increased correlation times for momentum fluctuations of the particles in the thermostat regions, as the size of the system increases. The scaling of $\langle J^2 \rangle_{0}$ with $L$ thus directly reports on slowly decaying heat current correlations\cite{lepri1998anomalous}.

From both the nonequilibrium and equilibrium calculations, we can explicitly compute $\kappa_L$, which is shown in Fig.~\ref{Fi:2} for the nanotubes and graphene sheets as functions of $L$. In order to place them on the same scale, we plot them relative to their values for $L=5$ nm, and find that these data are well described by the scaling function $f_d(L)L$. We find quantitative agreement between the conductivities computed from equilibrium fluctuations of the time integrated current, as well as direct nonequilibrium simulations, as expected from linear response and the Gaussianity of the current distributions in Fig.~\ref{Fi:1}. For both sets of systems, the conductivity initially increases linearly with $L$, signifying the role of ballistic phonon modes in transporting the energy for small systems. For $L>20$ nm, we find that $\kappa_L$ scales sub-linearly in a manner that depends on dimensionality.  This crossover length is consistent with estimations of the phonon mean free path using the speed of sound and heat capacity, and with extrapolations based on Matthiessen's rule\cite{salaway2014molecular}. This value is significantly lower other estimates of the phonon mean free path evaluated at low frequencies.\cite{saaskilahti2015frequency} It is also much smaller than mean free paths inferred from the Boltzmann equation based calculations of the thermal conductivity,\cite{fugallo2014thermal} however such calculations are not typically capable of describing anomalous transport due to their perturbative treatment of anharmonicities and neglect of nonMarkovian effects.\cite{alexanian2014classical}
We find that for the $1d$ nanotubes, the thermal conductivity continues to increase as $\sqrt{L}$ over the range of system sizes studied, and can fit $\alpha=0.5\pm 0.05$ to the data. For the $2d$ sheets, we find the thermal conductivity increases as  $\ln \, L$ over the system sizes studied.  This particular anomalous behavior is consistent with lattice models with interaction potentials with high symmetry\cite{delfini2007anomalous}.

In models like the FPUT chain and hard spheres on a line, the breakdown in Fourier's law has been interpreted as the emergence of a Levy walk process for energy transport\cite{cipriani2005anomalous,delfini2007energy,das2014numerical,liu2014anomalous}. Specifically, rather than a normal diffusion, it is proposed that quasiparticles transport energy via a stochastic process in which ballistic motions with random direction occur over time intervals, $\tau$, drawn from a power-law distribution, $\phi(\tau)\sim \tau^{- 2-\alpha}$, where $\alpha$ is the same anomalous exponent as in $\kappa_L$ \cite{dhar2013exact}. Such motions result in a mean squared displacement of the energy carriers that scales super-diffusively, $\sim t^{2-\alpha}$.  In the limit of an infinite closed system, the prediction of super-diffusive spreading of energy can be tested by following the decay of a localized perturbation. 

\begin{figure}[t]
\begin{center}
\includegraphics[width=8.5cm]{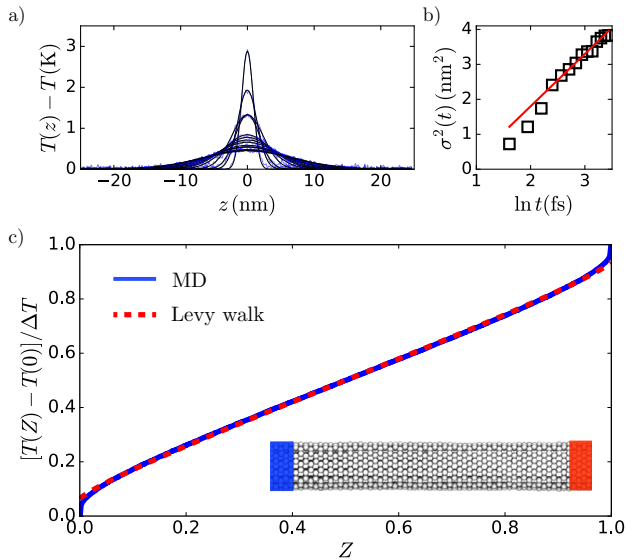}
\caption{Superdiffusion heat transport. a) Time dependent temperature profile following a localized perturbation. Blue lines are from molecular dynamics simulations, spaced 5 fs apart, and black line are approximate Gaussian fits to the Levy-stable distribution. b) Markers show the best fit variance from the time dependent temperature profiles as a a function of time, and the solid line is a fit to $\sigma^2(t)=a t^{3/2}+b$. c) Scaled temperature profile for $L=500$ nm nanotube solid line, and in the dashed line the best fit to Eq~\ref{Eq:Tofx}.}
\label{Fi:3}
\end{center} 
\end{figure}

We have considered an $L=$25 nm nanotube, with an initially localized temperature profile, $T(z)=T+T_1[ \Theta(L-\delta-z) -\Theta(L+\delta-z)]$, where $\Theta$ denotes a Heaviside function, and evolved in time in the absence of coupling to the thermostats. This is shown in Fig.~\ref{Fi:3}a), where we have taken $T_1=$700 K, and show the time dependent temperature profile averaged over 10$^4$ random initial conditions. By fitting the width of these distributions using a local Gaussian form, $T(z)=T+\Delta T(t) \exp[-(z-L)^2/2 \sigma^2(t)]$, we conclude that the kinetic energy spreads super-diffusively as predicted from the Levy walk model, $\sigma^2(t) \sim t^{2-\alpha}$, with an exponent consistent with $\alpha=1/2$, shown in Fig.~\ref{Fi:3}b). Scattering based imaging techniques like StroboSCAT\cite{delor2018imaging} could potentially be used to confirm this analogous behavior in suspended graphene or MoS$_2$ sheets. 

The Levy walk model also makes a prediction regarding the nonequilibrium steady state temperature profile for an open system\cite{lepri2011density}. In the limit that $L$ is large, for $\alpha=1/2$, the temperature profile is given by 
\begin{equation}
\label{Eq:Tofx}
T(Z) =T(0)+ \frac{\Delta T}{C} \int_0^Z dZ' \frac{1}{(Z'-Z'^2)^{1/4}}
\end{equation}
where $Z=z/L$ and $C=8 \Gamma(3/4) \Gamma(7/4) /3 \sqrt{\pi}$ where $\Gamma$ is the Gamma function\cite{dhar2013exact}. Shown in Fig.~\ref{Fi:3}c) is the scaled temperature profile for a $L=500$ nm carbon nanotube and $\Delta T=100$ K. The profile is nonlinear, as has been previously observed, with temperature jumps at the thermostat boundaries due to well documented Kapitza resistances\cite{pollack1969kapitza}. For large nanotubes, we find that the Kapitza resistance contribution shrinks, and the profiles converge to a stable nonlinear form, which can be well fit by Eq.~\ref{Eq:Tofx}, and deviates significantly from the linear profile expected from Fourier's law. This persistent nonlinearity is a consequence of the nonlocal relation between the heat current and local temperature gradients resulting from the Levy walk Green's function\cite{dhar2013exact}. The emergence of a Levy walk can be understood as a consequence of the confinement of momentum fluctuations that correlates motion over long times. This nonlocal transport mechanism can be utilized in nanophononic devices to enhance thermal rectification effects.\cite{yang2007thermal}

In this work, we have shown that carbon lattices described by a detailed molecular model exhibit anomalous heat transport over the range of system sizes studied, for both 1$d$ nanotubes and $2d$ graphene sheets. The anomalous exponent $\alpha$ that relates the divergence of the thermal conductivity to the system's characteristic length was extracted from direct nonequilibrium calculations in steady state and transiently, and from spontaneous equilibrium fluctuations, where the latter was enabled by recently developed importance sampling method. While the calculations are restricted to $L<2$ $\mu$m, all means of extracting $\alpha$ consistently find that $\kappa_L \sim L^{1/2}$ for 1$d$ carbon lattices, and $\kappa_L \sim \ln L$ for 2$d$ carbon lattices, and are consistent with the Levy walk model of low $d$ heat transport. While we have focused on carbon materials, out results should extend to other materials capable for being synthesized into nanotubes and sheets like boron nitride.

\emph{Acknowledgments}. The authors thank Garnet Chan and Kranthi Mandadapu for helpful discussions, the US National Science Foundation via grant CHE-1665333 and the UC Berkeley College of Chemistry for support. 

%


\end{document}